\documentclass[aps,prl,twocolumn,amsmath,amssymb,superscriptaddress]{revtex4}
\usepackage{graphicx}
\usepackage{amssymb}
\usepackage{natbib}
\usepackage{color}

\newcommand{\beq}{\begin{eqnarray}}
\newcommand{\eeq}{\end{eqnarray}}

\begin{document}

\title{Direct observation of magnon-phonon coupling in yttrium iron garnet}
\author{Haoran Man}
\thanks{These authors made equal contributions to this paper}
\affiliation{Department of Physics and Astronomy,
Rice University, Houston, Texas 77005, USA}
\author{Zhong Shi}
\thanks{These authors made equal contributions to this paper}
\affiliation{Department of Physics and Astronomy,
University of California, Riverside, California 92521, USA}
\affiliation{School of Physics Science and Engineering, Tongji University, Shanghai 200092, China}
\author{Guangyong Xu}
\affiliation{NIST Center for Neutron Research, National Institute of Standards and Technology, Gaithersburg, Maryland 20899, USA}
\author{Yadong Xu}
\affiliation{Department of Physics and Astronomy,
University of California, Riverside, California 92521, USA}
\author{Xi Chen}
\affiliation{Materials Science and Engineering Program, Texas Materials Institute, The University of Texas at Austin, Austin, Texas 78712, USA}
\author{Sean Sullivan}
\affiliation{Materials Science and Engineering Program, Texas Materials Institute, The University of Texas at Austin, Austin, Texas 78712, USA}
\author{Jianshi Zhou}
\affiliation{Materials Science and Engineering Program, Texas Materials Institute, The University of Texas at Austin, Austin, Texas 78712, USA}
\author{Ke Xia}
\affiliation{Department of Physics, Beijing Normal University, Beijing 100875, China}
\author{Jing Shi}
\email{jings@ucr.edu}
\affiliation{Department of Physics and Astronomy,
University of California, Riverside, California 92521, USA}
\author{Pengcheng Dai}
\email{pdai@rice.edu}
\affiliation{Department of Physics and Astronomy,
Rice University, Houston, Texas 77005, USA}
\affiliation{Department of Physics, Beijing Normal University, Beijing 100875, China}

\date{\today}

\begin{abstract}
The magnetic insulator yttrium iron garnet (YIG) with a ferrimagnetic transition temperature of $\sim$560 K has been widely used in microwave and spintronic devices.  Anomalous features in the spin Seeback effect (SSE) voltages have been observed 
in Pt/YIG and attributed to the magnon-phonon coupling.  Here we use inelastic neutron scattering to map out low-energy spin waves 
and acoustic phonons of YIG at 100 K as a function of increasing magnetic field.  By comparing the zero and 9.1 T data, we find
that instead of splitting and opening up gaps at the spin wave and acoustic phonon dispersion  
intersecting points, magnon-phonon coupling in YIG enhances the hybridized scattering intensity. These results are different from expectations of conventional spin-lattice coupling,
calling for new paradigms to understand the scattering process of magnon-phonon interactions and the resulting magnon-polarons.  
\end{abstract}

\maketitle
 
Spin waves (magnons) and phonons are propagating disturbance of the ordered magnetic moment and lattice vibrations, respectively.
They constitute two fundamental quasiparticles in a solid and can couple together to form a hybrid 
quasiparticle \cite{Heisenberg,Lovesey}. Since our current understandings of these quasiparticles are based on linearized models that ignore 
all the high-order terms than quadratic terms and neglect interactions among the quasiparticle themselves \cite{landau}, magnons and phonons   
are believed to be stable and unlikely to interact and breakdown for most purposes \cite{Zhitomirsky}. 
Therefore, discovering and understanding how the otherwise stable magnons and phonons  
can couple and interact with each other to influence the electronic properties of solids are one of the central themes in modern condensed 
matter physics.

In general, spin-lattice (magnon-phonon) coupling can modify magnon in two different ways.  First, the static lattice distortion 
induced by the magnetic order may affect the anisotropy of magnon 
exchange couplings, as seen in the spin waves of iron pnictides with large in-plane magnetic exchange anisotropy \cite{zhao09}.  
Second, the dynamic lattice vibrations interact with time-dependent spin waves may give rise to significant magnon-phonon coupling \cite{Kamra15,Shen15}.
 One possible consequence of such coupling is to create energy gaps in the magnon
dispersion at the nominal intersections of the magnon and phonon modes \cite{Anda76,Guerreiro15}, as seen in antiferromagnet (Y,Lu)MnO$_3$ \cite{oh16}. 
Alternatively, magnon-phonon coupling may give rise to spin-wave broadening at the magnon-phonon crossing points \cite{dai00}. In both cases, we expect the integrated intensity of hybridized excitations 
at the intersecting points to be the sum of separate magnon and phonon scattering intensity without spin-lattice coupling \cite{Anda76}. Finally, 
if magnon and phonon lifetime-broadening is smaller than their interaction strength, the resulting mixed quasiparticles can form  
magnon polarons \cite{Kamra15,Shen15}.

Here we use inelastic neutron scattering to study low-energy ferromagnetic magnons and acoustic phonons in  
 the ferrimagnetic insulator yttrium iron garnet (YIG) with chemical formula Y$_3$Fe$_5$O$_{12}$ [Figs. 1(a)-1(d)] \cite{Geller,Plant,Cherepanov93}.
At zero field and 100 K, we confirm the quadratic wave vector dependence of the magnon energy, $E=Dq^2$, where $D$ is the effective spin wave stiffness constant and $q$ is momentum transfer 
(in \AA$^{-1}$ or 10$^{10}$m$^{-1}$) away from a Bragg peak [Fig. 1(e)] \cite{Plant,Cherepanov93,Harris63,Srivastava87,barker,Xie17,Princep17}.  We also confirm the linear dispersion of the TA phonon mode  [Fig. 1(e)].  Upon application of a magnetic field $H_0$, a spin gap of the magnitude $g H_0$ ($g\approx 2$ is the 
Land$\rm \acute{e}$ electron spin $g$-factor) opens and lifts up  
spin waves spectra away from the field-independent phonon dispersion  
[Figs. 1(f) and 1(g)] \cite{Plant,Cherepanov93}.  By comparing the zero and 9.1 T field 
wave vector dependence of the spin wave spectra, we find
that instead of splitting and opening up gaps at the spin wave and acoustic phonon dispersion  
intersecting points, hybridized magnon polaron scattering  at the intersecting points has larger intensity
at zero field and magnons remain unchanged at other wave vectors as shown schematically in the bottom panels of Figs. 1(f) and 1(g). 
This is different from the expectations of conventional magnon-phonon interaction, where hybridized polaronic excitations
at the crossing points should have the sum of separate magnon and phonon scattering intensity, and become broader in energy due to the repulsive magnon-phonon dispersion curves \cite{Anda76,Guerreiro15,dai00,oh16}.  Our results thus reveal a new magnon-phonon coupling mechanism, calling for
a new paradigm to understand the scattering process of magnon-phonon interactions and the resulting magnon polarons \cite{Kikkawa16}.

\begin{figure}
\includegraphics[width = 8cm]{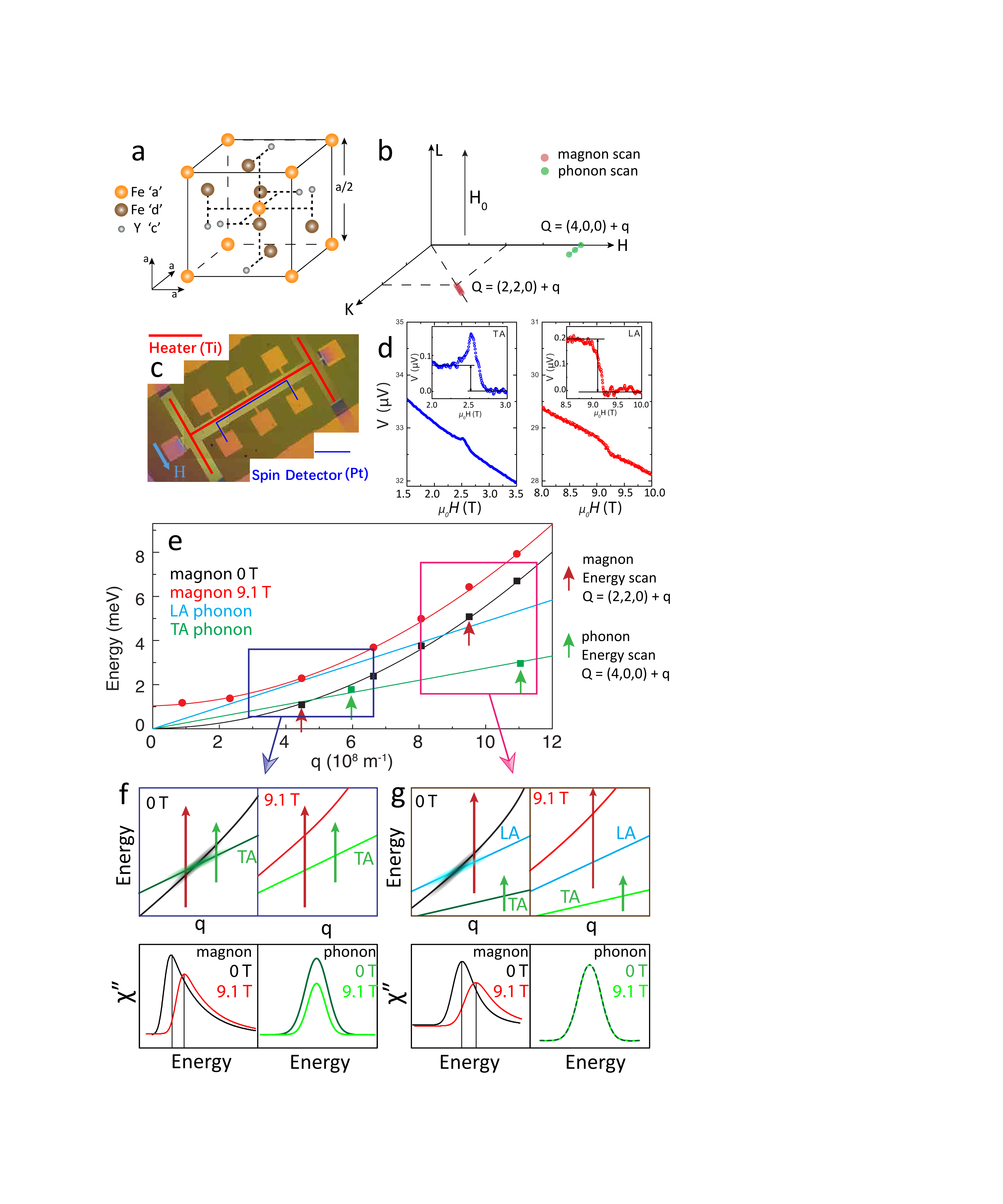}
\caption{(a) The full unit cell of YIG comprises eight cubes that are related by glide planes to the basic cube shown in the figure. 
(b) The corresponding reciprocal space with the $[H,K,0]$ scattering and vertical magnetic field $H_0$.  The red and green solid circles mark the positions of reciprocal space where we probe 
spin waves and acoustic phonons, respectively. (c) A picture of the Pt/YIG device used for SSE measurements. (d)  
SSE voltage in the field ranges where anomalous features appear at 100 K. 
(e) Magnon and phonon dispersions of YIG at 100 K and different magnetic fields. The black squares and solid red circles are data from 0 T and 9.1 T measurements, respectively.
The $q\approx 4.5\times 10^8$ m$^{-1}$ point corresponds to 
$\Delta Q=0.062$ in Fig. 2(d). The black and red solid
lines are quadratic ferromagnetic spin wave fit to the data. The blue and red boxes indicate magnon-phonon crossing points.  The green and blue solid lines are TA (with phonon velocity $C_\perp\approx 3.9\times 10^3$ m/s)and LA ($C_{||}\approx 7.2\times 10^3$ m/s)
phonons, respectively \cite{Kikkawa16}. (f,g) The expanded view of the blue and red boxes in (e), respectively. The bottom panels in (f,g) 
summarize the results obtained in our measurements on the magnetic field effect on 
spin waves, hybridized excitations, and TA phonons.   
}
\end{figure}

We chose to study magnon-phonon coupling in YIG because it is arguably 
the most important material used in microwave and recent spintronic devices \cite{Chumak15}. In addition to having a ferrimagnetic ordering temperature of $\sim$560 K suitable for room temperature applications, YIG can be grown with exceptional quality, and has the lowest Gilbert damping of any known materials and a narrow magnetic resonance linewidth allowing transmission of spin waves over macroscopic distances \cite{Serga10,Chumak14,Kajiwara10}. The spin Seebeck effect (SSE), which allows spin currents
produced by thermal gradients in magnetic materials to be transmitted and converted to charge voltages in a heavy metal such as Pt, is one of the most technologically relevant thermoelectric phenomena to be used in `spin caloritronic' devices \cite{Bauer12,Uchida08,Uchida10,Uchida10a,Jaworski11,Hoffmann15}. 
 In the case of a Pt film 
on the surface of a polished single-crystalline YIG slab (Pt/YIG)
 [Fig. 1(c)] \cite{Qiu13}, anomalous features in  magnetic field dependence of the SSE voltages at low temperatures are attributed to the magnon-phonon interaction 
at the ``touching'' points between the magnon and
transverse acoustic (TA) and longitudinal acoustic (LA) phonon as magnon dispersion curve is lifted by the applied field while phonon is not affected by the field
[Fig. 1(d)] \cite{Kikkawa16}. While we find no anomaly at the magnon and TA/LA acoustic phonon touching points, our data reveal clear evidence for magnon-phonon
interaction at zero field, consistent with the formation of magnon polarons.

\begin{figure}
\includegraphics[width = 8cm]{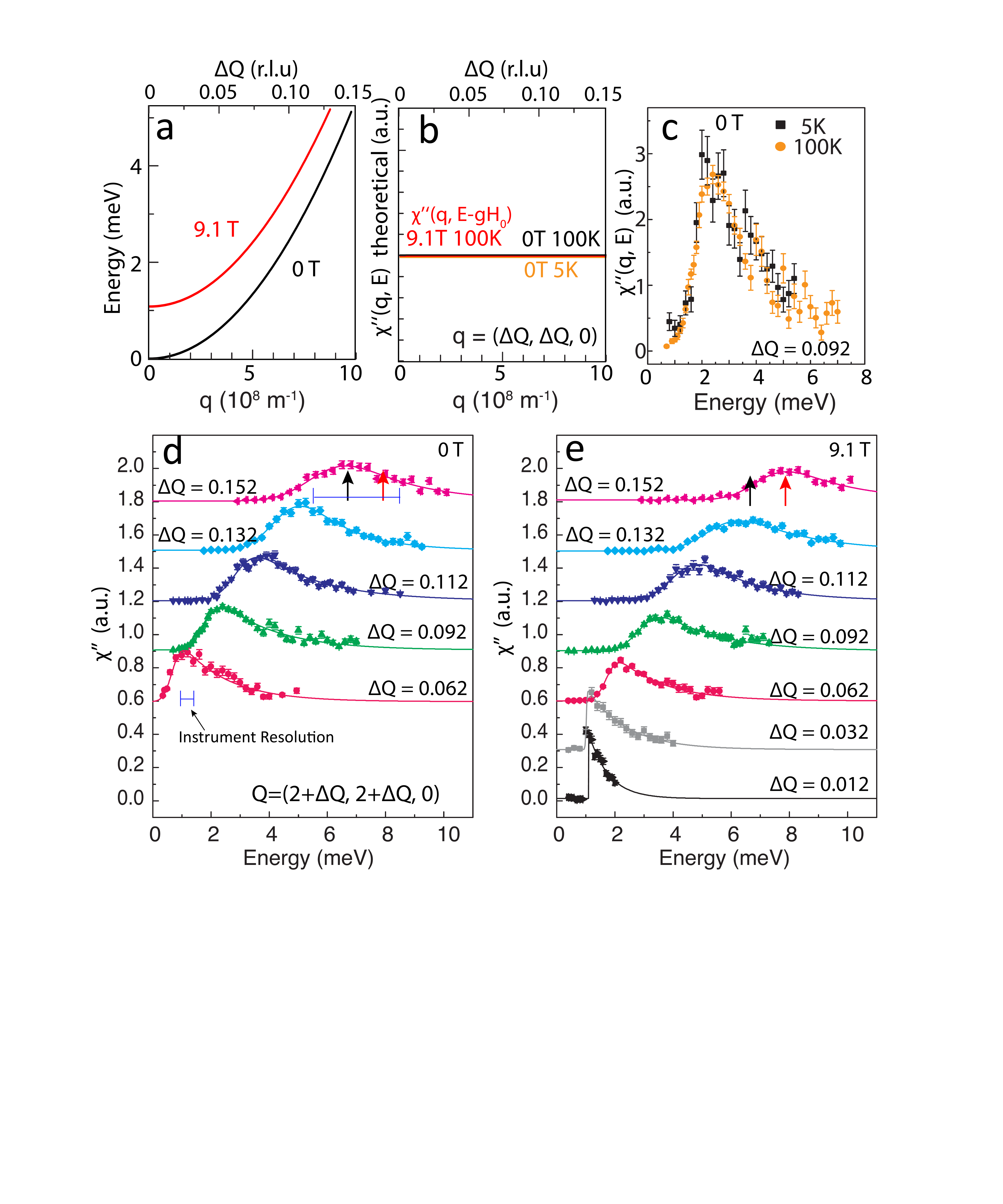}
\caption{(a) Schematic illustration of the expected magnon dispersions at 0 T and 9.1 T for a simple ferromagnet.
(b) The expected temperature, magnetic field dependence of low-energy $\chi^{\prime\prime}({\bf Q},E)$ for simple ferromagnet obtained from SpinW software package \cite{Toth}. Here the magnetic field induced spin gap $gH_0$ has been subtracted in the 9.1 T $\chi^{\prime\prime}(q,E-gH_0)$ (red).
The upper and bottom units are $\Delta Q$ and $q$, respectively. 
(c) Our estimated $\chi^{\prime\prime}({\bf Q},E)$ with ${\bf Q}=(2.092,2.092,0)$ at 5 K and 100 K after correcting measured $S({\bf Q},E)$ 
for the background and Bose-population factor. 
(d,e) The estimated $\chi^{\prime\prime}({\bf Q},E)$ at 0 T and 9.1 T, respectively, after correcting for background and Bose population factor.
 Scans at different wave vectors are lifted up by 0.3 sequentially. The black and red arrows marks the peak positions at 0 T and 9.1 T, respectively.}
\end{figure}

Our neutron scattering experiment was carried out at NIST center for neutron research, Gaithersburg, Maryland \cite{SI}. 
The full body-centered-cubic unit cell of YIG with space group $Ia3d$ comprises eight cubes that are related by glide planes 
to the basic cube as shown in Fig. 1(a),
where the metallic atomic sites are labelled as `a', `d', and `c' \cite{Plant}. Using the cubic lattice parameter of $a=b=c=12.376$ \AA,  
we define momentum transfer ${\bf Q}$ in three-dimensional (3D) reciprocal space in \AA$^{-1}$ as $\textbf{Q}=H\textbf{a}^\ast+K\textbf{b}^\ast+L\textbf{c}^\ast$, where $H$, $K$, and $L$ are Miller indices and
${\bf a}^\ast=\hat{{\bf a}}2\pi/a$, ${\bf b}^\ast=\hat{{\bf b}}2\pi/a$, ${\bf c}^\ast=\hat{{\bf c}}2\pi/a$ [Figs. 1(a) and 1(b)]. 
Consistent with Ref. \cite{Kikkawa16}, the magnetic field dependence of SSE voltage on our Pt film on YIG contains two anomalous features at 2.5 T and 9.1 T [Figs. 1(c)-1(e)] \cite{SI,Jiang16,Xu14,SMWu15,Collet17}.

The sample for neutron scattering experiments was oriented with 
 $a$ and $b$($a$)-axis of the crystal in the horizontal $[H,K,0]$ scattering plane [Fig. 1(b)] and mounted inside a 10 T vertical field magnet. 
In this geometry, we measured magnon dispersion around $(2, 2, 0)$ and phonon dispersion around $(4, 0, 0)$.
 The momentum transfers $\textbf{Q}$ at these wave vectors are $\textbf{Q}_{magnon}=(2+\Delta Q, 2+\Delta Q, 0)$ and $\textbf{Q}_{phonon}=(4,\Delta Q,0)$ for TA phonon [Fig. 1(b)].
For convenience, we calculate relative momentum transfer as $q=2\pi\sqrt{2}\Delta Q/a$ for magnon and $q=2\pi\Delta Q/a$ for phonon. 
We chose $(2,2,0)$ for magnetic and $(4,0,0)$ for phonon measurements because of their huge differences in nuclear structure factors [4.75 at $(2,2,0)$ versus 50.5 at $(4,0,0)$],
 which is directly related to the acoustic phonon intensity. Although we expect to find mostly magnetic scattering at $(2,2,0)$ and phonon scattering at $(4,0,0)$, the finite Fe$^{3+}$ magnetic form factor 
of $\left|F({\bf Q})\right|$ means that there are still magnetic contributions to the phonon scattering at $(4,0,0)$ ($\left|F(2,2,0)\right|^2/\left|F(4,0,0)\right|^2\approx 1.86$).

\begin{figure}
\includegraphics[width = 8cm]{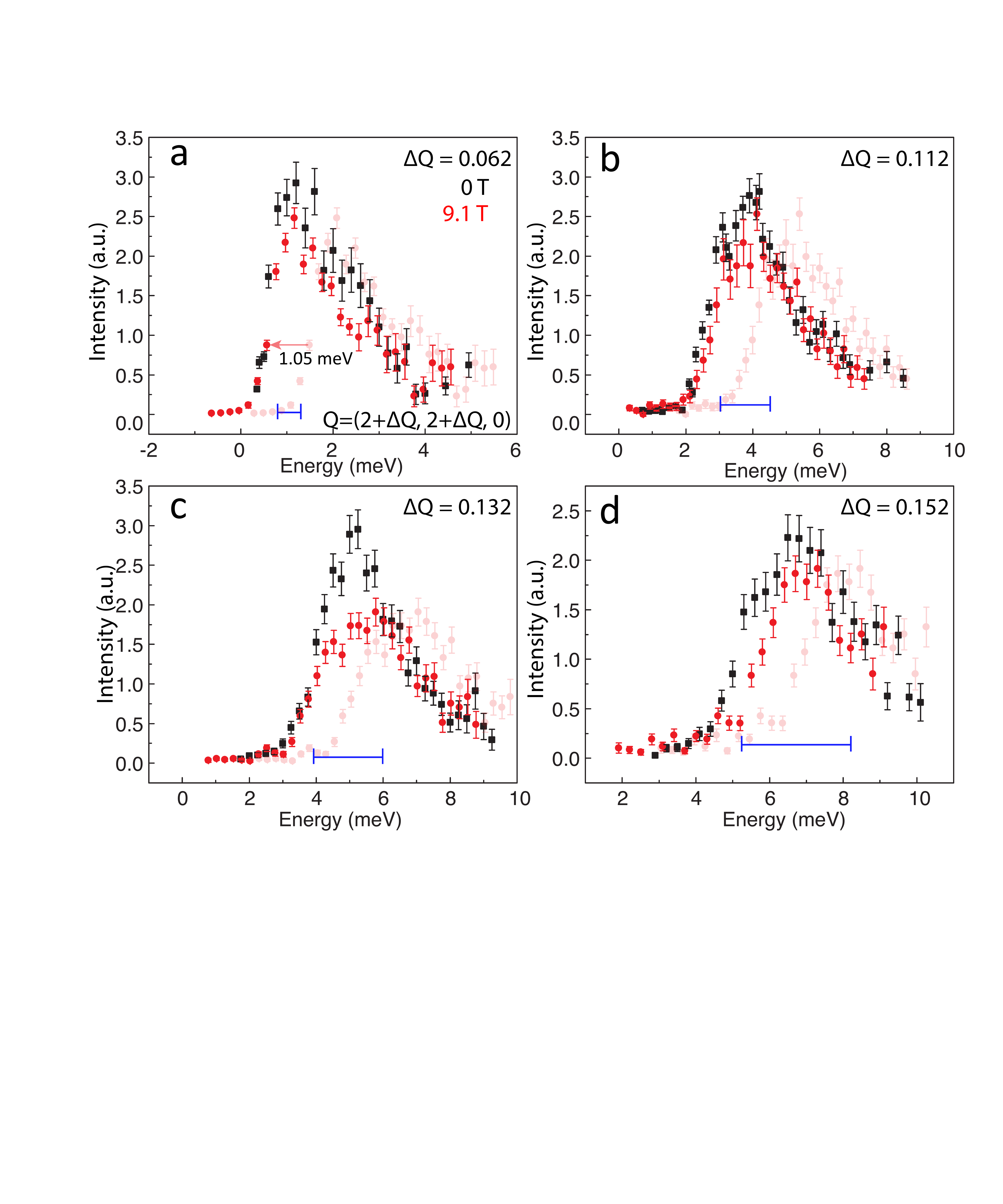}
\caption{(a,b,c,d) Comparison of the estimated $\chi^{\prime\prime}({\bf Q},E)$ as a function of increasing wave vector at 0 T (black) and 9.1 T (red).  The  
 9.1T data is shifted by 1.05 meV to accommodate the field induced energy shift. Light red dots represents the original data position of the 9.1 T data. The horizontal bars 
are estimated instrumental energy resolution based on magnon dispersion at 100 K.   
}
\end{figure}

Magnetic neutron scattering directly measures the magnetic
scattering function $S({\bf Q},E)$, which is proportional
to the imaginary part of the dynamic susceptibility $\chi^{\prime\prime}({\bf Q},E)$ through
$S({\bf Q},E)\propto \left|F({\bf Q})\right|^2\chi^{\prime\prime}({\bf Q},E)/[1-\text{exp}(-\frac{E}{k_BT})]$, where 
$E$ is the magnon energy, $k_B$ is the Boltzmann constant \cite{Lovesey}. 
Although YIG is a ferrimagnet, its low-energy spin waves can be well described as a simple ferromagnet \cite{barker}. 
In the hydrodynamic limit of long wavelength (small-$q$) and small energies, we expect $E=\Delta_0+gH_0+Dq^2$ for spin wave dispersion, where $\Delta_0$ is the possible intrinsic spin anisotropy gap, $gH_0$ is the size of
the magnetic field induced spin gap, and $D$ is in units of meV\AA$^{2}$ [Fig. 2(a)] \cite{Plant,Cherepanov93,Harris63,Srivastava87}. In addition, 
for a pure magnetic ordered system without spin-lattice interaction, 
we expect that $\chi^{\prime\prime}({\bf Q},E)$ to be 
independent of temperature at temperatures well below the magnetic ordering temperature 
and applied magnetic field after correcting for the field-induced spin gap $gH_0$ [Fig. 2(b)] \cite{dai,Ye07,Toth}.

\begin{figure}
\includegraphics[width = 8cm]{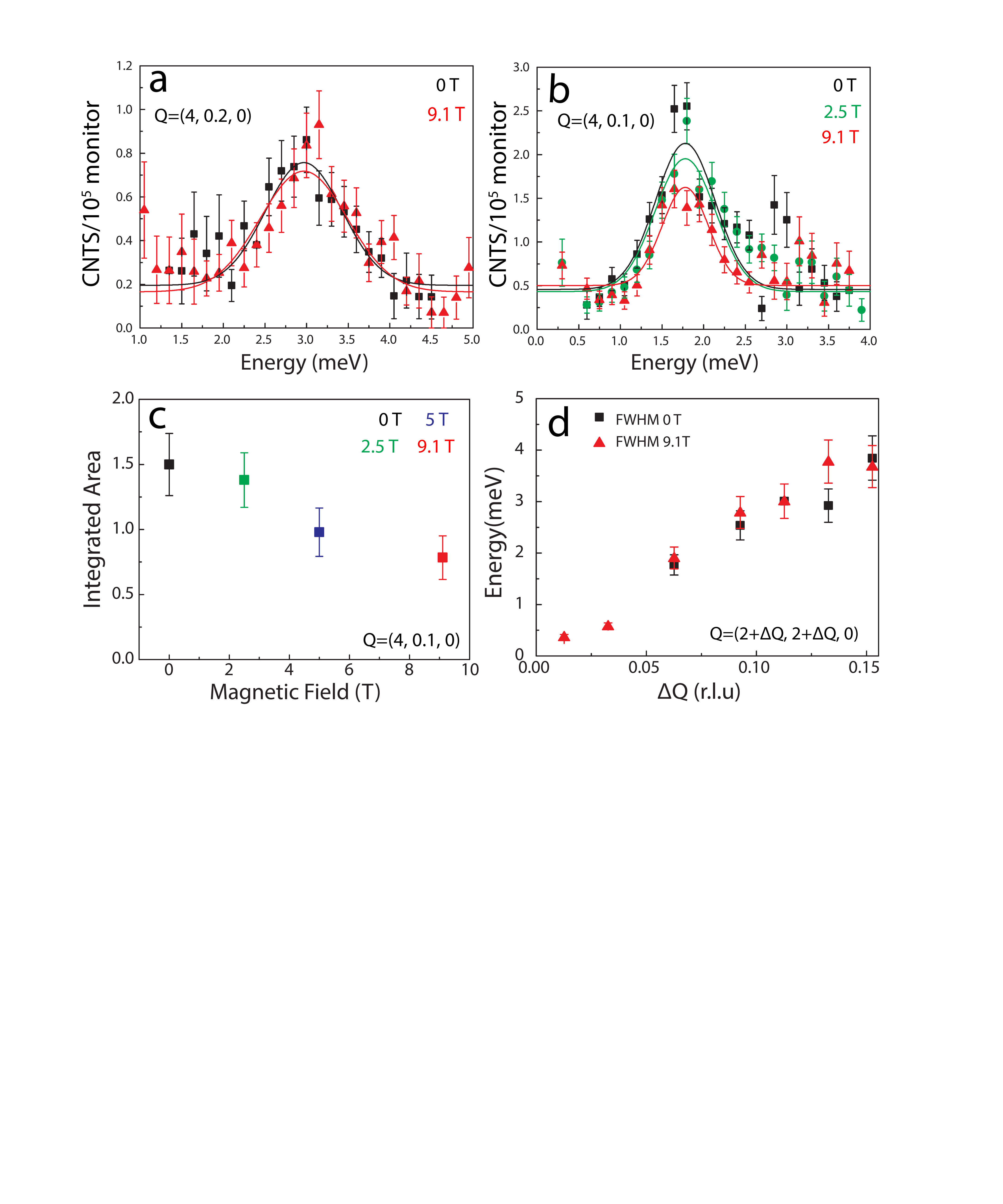}
\caption{(a) Energy scan of $S({\bf Q},E)$ at ${\bf Q}=(4,0.2,0)$, a position far away from magnon-phonon crossing points, and 100 K to probe TA phonon at 0 T and 9.1 T
(b) Energy scan of $S({\bf Q},E)$ to probe magnon-phonon hybridized excitations at ${\bf Q}=(4, 0.1, 0)$ near
magnon-phonon crossing point at 0 T, 2.5 T, and 9.1 T. (c) Magnetic field dependence of the integrated intensity of 
magnon-phonon hybridized excitations at 100 K and ${\bf Q}=(4, 0.1, 0)$. (d) FWHM of the magnon at 0 T and 9.1 T as a function of
$\Delta Q$.
 }
\end{figure}

To determine if temperature and magnetic field dependence of spin waves in YIG follow these expectations, we measured wave vector dependence of magnon energy of YIG at different temperatures and
magnetic fields. Figure 2(c) shows our estimated constant-${\bf Q}$ scans [${\bf Q}=(2.092,2.092,0)$ or $\Delta Q=0.092$ rlu]
 of $\chi^{\prime\prime}({\bf Q},E)$ at 5 K (filled black squares) and 100 K (filled orange circles).  Consistent with the expectation, we see that $\chi^{\prime\prime}({\bf Q},E)$ at these two temperatures are identical within the errors of the measurement. Figure 2(d) shows 
constant-${\bf Q}$ scans of spin waves of YIG at 100 K and 0 T.  At ${\bf Q}=(2.062,2.062,0)$ or $\Delta Q=0.062$, $\chi^{\prime\prime}({\bf Q},E)$ has a clear peak in energy 
that is slightly larger than
the instrumentation resolution (horizontal bar).  With increasing $\Delta Q$, the peak in $\chi^{\prime\prime}({\bf Q},E)$ moves progressively to higher energies.
We have attempted but failed to fit the spin wave spectra with a simple harmonic
oscillator generally used for a ferromagnet \cite{Ye07}. This may be consistent with recent inelastic neutron scattering study of YIG that reveals
the need to use long range magnetic exchange couplings to fit the overall spin wave spectra \cite{Princep17}. 
By fitting the spin wave spectra at zero field with an exponentially modified Gaussian peak function \cite{SI}, we obtain the magnon dispersion curve as shown in Fig. 1(e). Fitting the dispersion curve with $E=\Delta_0+Dq^2$ yields 
$\Delta_0\approx 0$ and $D=580\pm 60$ meV\AA$^{2}$, consistent with earlier work giving $D\approx 533$ meV\AA$^2$ \cite{Cherepanov93}.

Upon application of a 9.1 T field at 100 K, we expect the magnon dispersion curve to be lifted by $gH_0\approx 1$ meV.  This would be consistent with the observation of a sharp gap below 1.05 meV 
in constant-${\bf Q}$ scan at ${\bf Q}=(2.012,2.012,0)$ ($\Delta Q=0.012$) [Fig. 2(e)]. Constant-${\bf Q}$ scan at ${\bf Q}=(2.032,2.032,0)$ 
shows similar behavior. Figure 2(e) also shows constant-${\bf Q}$ scans at identical wave vectors as those in Fig. 2(d) at 0 T.  
Using data in Fig. 2(e), we plot the magnon dispersion at 9.1 T field in Fig. 1(e).
Consistent with the expectation, we see a clear $gH_0$ upward shift in magnon energy but the spin wave stiffness $D$ remains unchanged.

To quantitatively determine the magnetic field effect on $\chi^{\prime\prime}({\bf Q},E)$ of YIG, we compare $\chi^{\prime\prime}({\bf Q},E)$ at 0 T with those at 9.1 T.
Figure 3(a)-3(d) summarizes the energy dependence of $\chi^{\prime\prime}({\bf Q},E)$ after down shifting the 9.1 T data by $gH_0=1.05$ meV.
At $\Delta Q=0.062$, the scan along the red arrow direction near the magnon-phonon crossing point as shown in Fig. 1(f), we see that  $\chi^{\prime\prime}({\bf Q},E)$ at 9.1 T field is lower 
in intensity compared with those at 0 T.  On moving to $\Delta Q=0.10$ with no magnon-phonon crossing, $\chi^{\prime\prime}({\bf Q},E)$ at 0 T and 9.1 T are virtually identical as expected.
At the second magnon-phonon crossing point with $\Delta Q\approx 0.13$ [see red arrow in Fig. 1(g)], the differences between  $\chi^{\prime\prime}({\bf Q},E)$ at 0 T 
and 9.1 T are even more obvious, with intensity at 0 T considerably larger than that at 9.1 T [Fig. 3(c)]. Finally, on moving to $\Delta Q=0.152$ well above the 
magnon-phonon crossing point wave vectors [Fig. 1(e)], we again see no obvious difference in $\chi^{\prime\prime}({\bf Q},E)$ between 0 T and 9.1 T.

Figure 3 shows that magnetic field dependence of $\chi^{\prime\prime}({\bf Q},E)$ is highly wave vector selective, revealing clear magnetic field induced
intensity reduction in $\chi^{\prime\prime}({\bf Q},E)$ at wave vectors associated with magnon-phonon crossing points
 while having no effect at other wave vectors.  To confirm the presence of TA phonon and determine its magnetic field effect, we carried out TA phonon measurements near $(4,0,0)$, which has 
a rather large nuclear structure factor compared with $(2,2,0)$. Figure 4(a) shows energy scans of at ${\bf Q}=(4,0.2,0)$ and 100 K, which is along the green arrow direction in Fig. 1(g) 
and far away from the magnon dispersion.  The spectra reveal a clear magnetic field independent peak at $E\approx 3$ meV, confirming the TA phonon nature 
of the scattering.  Figure 4(b) shows similar energy scan at ${\bf Q}=(4,0.1,0)$ and 100 K, which is 
along the green arrow direction and 
near the magnon-phonon crossing point in Fig. 1(f). 
At 0 T, we see a peak around $E\approx 1.7$ meV consistent with dispersions of magnon and TA phonon.
With increasing field to 2.5 T and 9.1 T, the intensity of the peak decreases, but its position in energy remains unchanged [Fig. 4(b)].
Figure 4(c) shows magnetic field dependence of the integrated intensity, confirming the results in Fig. 4(b). 
Since the energy of the magnon should increase with increasing magnetic field, the field independent nature of the peak position 
in Fig. 4(b) suggests that the mode cannot be a simple addition of magnon and phonon, but most likely arises from hybridized magnon polarons \cite{Kamra15,Shen15}.
Figure 4(d) shows the full width at half maximum (FWHM) of the magnon width at 0 T and 9.1 T. Within the errors of our measurements, we see no energy width
change in the measured wave vector region.

Our results provided compelling evidence for the presence of magnon-phonon coupling in YIG at the magnon-phonon crossing points at zero field.  This is clearly different from the SSE measurements, where anomalies are only seen at the critical
fields that obey ``touch'' condition at which the mangnon energy and group velocity agree with that of the TA/LA phonons.
 When the applied field is less than the critical field, the magnon dispersion 
has two intersections with TA/LA phonon modes.  When the applied field is larger than the critical field, the magnon dispersoin is separated from 
the TA/LA phonon modes.  In the theory of hybrid magnon-phonon excitations \cite{Kamra15,Shen15}, the SSE anomalies occur at magnetic fields 
and wave vectors at which the phonon dispersion curves are tangents to the magnon dispersion, where the effects of the 
magnon-phonon coupling are maximized \cite{Flebus17}. While our findings of a novel magnon-phonon coupling at zero field 
are consistent with the formation of magnon-polarons in YIG \cite{Kamra15,Shen15}, they are not direct proof that magnon-polaron formation alone causes anomalous
features in the magnetic field and temperature dependence of the SSE. Other effects, such as spin diffusion length, acoustic quality of the YIG film, and magnon spin conductivity also play an important role in determining the SSE anomaly \cite{Cornelissen17}.  Regardless of the microscopic origin of the SSE anomaly, our discovery suggests the need to understand why magnon-phononok interaction and the resulting magnon polarons enhance the hybridized excitations 
at the magnon-phonon intersection points.

The neutron scattering work at Rice is supported by the
U.S. DOE, BES DE-SC0012311 (P.D.).   The
materials work at Rice is supported by the Robert
A. Welch Foundation Grant No. C-1839 (P.D.). The work at UCR (J.S. and Z.S.) is supported as part of the SHINES, an Energy Frontier Research Center
funded by the U.S. DOE, BES under Award No. SC0012670.
YIG crystal growth at UT-Austin is supported by the Army Research Office MURI award W911NF-14-1-0016.

\end{document}